\documentstyle[12pt]{article}
\begin{document}
\begin{flushright}
OCHA-PP-180
\end{flushright}
\baselineskip 0.8cm
\begin{center}
 {\LARGE\bf Old Fashioned Duality for D-brane and String}\\
\bigskip
Rika ENDO, Koichi SEO$^\dagger$ and Akio SUGAMOTO\\
\medskip
{\it Department of Physics, Ochanomizu University \\
2-1-1, Otsuka, Bunkyo-ku, Tokyo 112-8610,  Japan\\
and\\
$\dagger$ Gifu City Women's College\\
7-1 Hitoichiba-kita-machi,  Gifu 501-0192, Japan }
 \end{center}    

\bigskip

\begin{abstract}
Old fashioned duality used to derive  the closed string field theory for
magnetic vortex from the gauge theory with Higgs scalar, is applied to the
string theories. The bosonic sring theory coupled to the Kalb-Ramond 2-form
field is dually transformed to the 6-brane theory coupled to the 7-form
field.  The old fashioned dual transformation is also examined for the Type
IIA and IIB superstrings.  For this  study, the string field theoretical
treatment of the bosonic and fermionic strings is developped based on the
reparametrization invariant formulation of strings by Marshall and Ramond. 
In order for the self-duality of the Type IIB superstring to appear, the
following dual correspondence may happen: the dual transformation of the
NS-NS field functional is the bosonization of the R-R one.  
\end{abstract}
\section{Introduction}
In the late 70's, we were in the first time excitement on the duality, in
which the quark confinement mechanism was searched based on the
Mandelstam-'t Hooft duality~\cite{'t Hooft Mandelstam}.  Many works were
carried out in this direction, using the so-called ``dual(-ity)
transformation"~\cite{Savit}, \cite{Abelian D-T}, \cite{non-Abelian D-T}. For
example, the Abelian Higgs model is ``dually transformed" into the
relativistic hydrodynamics of Kalb-Ramond and of Nambu ~\cite{K-R} coupled
to the vorticity source.  The latter model is further transformed to a
closed string field theory of the magnetic vortex coupled to the
anti-symmetric tensor fields called Kalb-Ramond fields, or the gauge field
of strings~\cite{Abelian D-T}.  In the course of the dual transformation
the coupling constant $g$ is naturally replaced by its inverse $1/g$.

Recently, D-brane has been introduced as a source of the closed strings,
and is understood as a soliton in the superstring theories, like a monopole
or a magnetic string in the gauge theories~\cite{D-brane}.   Since then,
the D-brane dynamics has begun to reveal the strong coupling regime of the
superstring theories, generating the second time excitement of the string
duality~\cite{string duality}.

The first and the second time dualities are quite analogous: Starting from
a gauge theory having the minimal coupling for an electric charge, another
theory is derived by the dual transformation, which has the minimal
coupling for the magnetically charged soliton of the former theory. 
Therefore, it is interesting to study the D-brane, a key element of  the
second time duality,  based on the old fashioned duality of the first time.
The relation between the electrically charged scalar field theory and the
string field theory of magnetic vortices is analogous to the relation
between the fundamental string theory and  the D-brane theory.

More recently, a duality is conjectured by Mardacena between the gauge
theories on the 4 dimensional boundary and the string theory (or the
supergravity theory as a low energy effective theory of the string theory)
on the 10 dimensional bulk~\cite{Maldacena}.  The example mentioned above,
giving the old-fashioned duality between the Abelian Higgs model and the
closed string field theory, is quite interesting, since the dual
relationship between  the gauge theory and the closed string theory is
quite analogous to the Mardacena's one.  In the Mardacena's duality
exisitence of the extra 5th dimension is essential, which represents the
energy scale of the thoery.  In the old-fashioned duality, the role of the
5th dimension is played in a sense by the vacuum expectation value of the
Higgs scalar $|\phi|$. We do not have yet the constructive way of finding
the dual partner of a given realisitic theory, not having conformal
symmetry nor supersymmetry.  In this respect, the old fashioned duality is
worth re-examining, from the present insights.

Preliminary stage of this work can be found in the lecture given at
Kashikojima Summer Institute by one of the authors
(A.S.)~\cite{Kashikojima}, and a part of this work has been studied in the
phD thesis of another one of the authors (R.E)~\cite{Endo phD}. 

 In this paper, the fundamentals such as the deformation and the sum of
curves, {\it etc.} are refined more rigorously, and the duality between the
fundamental string and the D-brane is clarified.  Furthermore, the old
fashioned dual transformation is examined for the superstrings of
Neveu-Schwarz and of Ramond, based on the field theoretical formulation of
the theories by Marshall and Ramond~\cite{Marshall-Ramond}.

\section{Old Fashioned Dual Transformation}
Dual transformation is a kind of  Fourier transformation performed in the
integrand of the partition function.  It is similar to the picture changing
transformation from x-representation to  p-representation in quantum
mechanics, where commutation relations
\begin{equation}
 [\hat{x},\hat{p}]=i, ~~~\mbox{and}~~~[\hat{x}, \hat{x}]=[\hat{p}, \hat{p}]=0.
\end{equation}
hold and its physical manifestation is the uncertainty relation  $(\Delta
p)^2 \cdot (\Delta x)^2 \geq 1$ .  Similarly, dual transformation from
${\bf E}$-representation to ${\bf B}$-representation gives 't Hooft algebra
which is given for the SU(N) gauge theory by
\begin{equation}
\hat{A}[C]\hat{B}[C']= \hat{B}[C']\hat{A}[C]\times e^{2\pi i L(C, C')/N},
~~~\mbox{and}  ~~~[\hat{A}[C], \hat{A}[C']]=[\hat{B}[C], \hat{B}[C']]=0,
\end{equation}
where $L(C, C')$ is the linking number of two closed curves $C$ and $C'$. 
Its uncertainty relation $(\Delta{\bf E})^2 \cdot (\Delta{\bf B})^2 \geq 1
$ gives a physical picture of the duality, that is, the squeezing of the
electric flux and that of the magnetic flux cannot be observed at the same
time. 

 In order to change ${\bf E}$ -representation to ${\bf B}$ -representation,
we simply perform the following dual transformation, a Fourier
transformation with respect to the field strengths performed in the
integrand of the partition function~\cite{Abelian D-T}\cite{non-Abelian
D-T}:
\begin{eqnarray}
&&\exp \left\{ i \int d^4x -\frac{\varepsilon}{4} F^{a}_{\mu\nu}F^{a
\mu\nu} \right\}  \nonumber  \hfill \\ 
&\propto&\int {\cal D}W^{a}_{\mu\nu}(x) \exp \left\{ i \int d^4x \left[
\frac{1}{4\varepsilon} W^{a}_{\mu\nu}W^{a\mu\nu}- \frac{1}{2}
\tilde{W}^{a}_{\mu\nu}F^{a\mu\nu} \right] \right\} ,\label{D-T}
\end{eqnarray}
where $\varepsilon$ is a dielectric constant,  $F^{a}_{\mu\nu}$ is the
field strength of the gauge field $A^{a}_{\mu}$, and $W^{a}_{\mu\nu}(x)$ is
an anti-symmetric tensor field (Kalb-Ramond field), which  becomes a
velocity potential of the hydrodynamics after the  dual transformation is
performed.

 Following Ref.~\cite{Abelian D-T}, we start with the action of the Abelian Higgs model,
\begin{equation}
S = \int d^{4}x \biggl[ 
-\frac{1}{4} F_{\mu\nu}F^{\mu\nu} + | (\partial_{\mu}+ i e A_{\mu}) \phi
|^2 - V(|\phi|^2) \biggr].
\end{equation}

If the above dual transformation in Eq.(\ref{D-T}) is carried out in the
partition function of the Abelian Higgs model and the subsequent
integration over $A_{\mu}$ is performed, the following dual action is
obtained:
\begin{eqnarray}
S^{\star}&=& \int d^{4}x \biggl[
-\frac{1}{2 e^2 |\phi|^2} \frac{1}{2} (V_{\mu})^2 - \frac{1}{4}
(W_{\mu\nu})^2 + (\partial_{\mu}|\phi|)^2 - V(|\phi|)  \nonumber \\
& & + \frac{1}{2} \frac{2 \pi}{e} W^{\mu\nu} \times \frac{1}{4
\pi}\epsilon_{\mu\nu\lambda\rho}(\partial^{\lambda} \partial^{\rho} -  
\partial^{\rho}\partial^{\lambda}) \chi(x) \biggr]. 
\end{eqnarray}
Here, $V_{\mu} = \partial^{\nu} \tilde{W}_{\nu\mu}$ is the velocity field
of the fluid, satisfying the continuity equation,
$\partial^{\mu}V_{\mu}=0$, and $\phi = |\phi| \exp(i \chi) $.  This model
gives a kind of ``relativistic hydrodynamics" which has been originally
formulated by Kalb and Ramond and by Nambu~\cite{K-R}. The singularity of
the Higgs' phase $\chi$ (that is the point around which the phase $\chi$
changes by $2\pi$ times interger n) gives a vorticity source,
\begin{equation}
\omega^{\mu, \nu} =
\frac{1}{4\pi}
\epsilon^{\mu\nu\lambda\rho}(\partial_{\lambda} \partial_{\rho} 
-\partial_{\rho}\partial_{\lambda}) \chi(x) = n \int d\tau d\sigma
\frac{\partial(y^{\mu}, y^{\nu})}{\partial(\tau, \sigma)}
\delta^{(4)}(x-y(\tau, \sigma)).
\end{equation} 
Here $x^{\mu}=y^{\mu}(\tau, \sigma)$ determines the world sheet swept by a
magnetic vortex, on which the $\chi$ field becomes singular, and the
integer $n$ represents the quantized vortex number. Recently this kind of
correspondence is rediscovered, and is applied to the study of the motion
of vortices in the superfluidity~\cite{superfluidity}. 

In this manner, the gauge coupling of the world sheet of the vortex
$y^{\mu}(\tau, \sigma)$ with the Kalb-Ramond field $W_{\mu\nu}$ is derived,
and the contribution of this interaction in the partition function reads
\begin{equation}
\exp \left( i \int_S d\sigma^{\mu\nu} \frac{1}{2} \frac{2\pi}{e} W_{\mu\nu}
\right) .\label{KR interaction}
\end{equation}

The above method is a general one of introducing a vortex (a ``bare
soliton") and of obtaining its coupling to the ``gauge field" (Kalb-Ramond
field in this case) with a strength $2\pi/e$, so that the dual model
obtained is suitable for the strong coupling expansion.
Therefore, this method may be useful to introduce D-branes and to obtain
their coupling to the Kalb-Ramond fields, since D-branes are the ``bare
solitons" of the string theory.  We will discuss this issue in the next
section. 

The contribution to the partition function of the classically charged
particle interacting with the electromagnetic field is $\exp(i e \int_C
dx^{\mu} A_{\mu}(x))$.  Furthermore, the field theory of charged particles,
that is the charged Higgs model, is known to be derived, by summing up all
the possible classical configurations having any number of closed circles
depicted by the charged particles. 
Similarly, summing up all the possible classical shapes of the world sheets
of the vortices, the field theory of string is obtained, since the coupling
of vortex and Kalb-Ramond field given in Eq. (\ref{KR interaction}) is the
same minimal cupling as the electromagnetic one.  Namely, we can arrive at
the following action of the closed string field theory, which is dually
related to the Higgs model~\cite{Abelian D-T}: 
\begin{equation}
S^{\star}=\sum_{C} \left\{ -\frac{1}{ \oint_{C}dx_{t}} \oint_{C} dx_t
\biggl| \left( \frac{\delta}{\delta\sigma^{\mu t}} - i \frac{2\pi}{e}
W_{\mu t} \right) \Phi[C] \biggr|^{2} - M_{0}^{2} \biggl| \Phi[C]
\biggr|^{2} \right\}, \label{K-R string action}
\end{equation}
where $C$ is a closed curve on which string field $\Phi[C]$ is defined, and
$M_{0}$ is a constant  including the entropy effect.  Once we have the
quantum field theory of the string, the probability for a closed string to
take the shape $C$ is given by $|\Phi[C]|^{2}$ as usual.

Furthermore, if we add the fermionic action $S_{f}$ to the Higgs model:
\begin{equation}
S_{f} = \int d^{4}x \biggl[
\bar{\psi} (i {\ooalign{\hfil/\hfil\crcr$\nabla$}} - m_{f} ) \psi + e
\bar{\psi} \gamma_{\mu} \psi A^{\mu} \biggr],
\end{equation}
then the corresponding dual action has the following additional term:
\begin{equation}
S_{f}^{\star} = \int d^{4}x \biggl[
\bar{\psi} (i {\ooalign{\hfil/\hfil\crcr$\nabla$}} - m_{f} ) \psi - n
\bar{\psi} \gamma_{\mu} \psi \partial^{\mu} \theta + \frac{1}{2} \frac{
2\pi}{e} W^{\mu \nu} \omega^{F}_{\mu \nu} - \frac{1}{4}
\frac{1}{|\phi|^{2}} (\bar{\psi} \gamma_{\mu} \psi )
(\bar{\psi} \gamma^{\mu} \psi ) \biggr],
\end{equation}
where the additonal vorticity appears from the fermion, 
\begin{equation}
\omega^{F}_{\mu \nu} = \frac{1}{4\pi} \epsilon_{\mu\nu\lambda\rho}
\partial^{\lambda} \left( \frac{1}{|\phi|^{2}} (\bar{\psi} \gamma^{\rho}
\psi ) \right).
\end{equation}

\section{Application of Old Fashioned Duality for D-branes}
In this section, we start with the closed string field theory coupled to
the Kalb-Ramond field (2-form field) which is a kind of the hydrodynamics
obtained by the dual transformation from the Abelian Higgs model in the
previous section.  Here, we take the superstringy space-time dimension of
$D=10$. 

 Our starting action is tentatively 
\begin{equation}
S^{(0)} = \sum_{C} \sum_{x(\in C)} \bigg|  \left(
\frac{\delta}{\delta\sigma^{t\mu} (x) } - i g A_{t\mu}(x) \right) \Phi[C]
\bigg|^2 + \sum_{x} - \frac{\varepsilon}{2 \cdot 3! } F_{\mu\nu\lambda}
F^{\mu\nu\lambda} - V(|\Phi[C]|^2) , 
\end{equation}
where $F_{\mu\nu\lambda}= \partial_{\mu}A_{\nu\lambda}+
\partial_{\nu}A_{\lambda\mu} + \partial_{\lambda}A_{\mu\nu}$.  We consider
the massless Kalb-Ramond field, so that the model has the gauge symmetry of
the string theory ({\it i.e.} Kalb-Ramond symmetry):
\begin{eqnarray}
\Phi[C] &\rightarrow& \Phi'[C] = e^{i \oint_{C} dx^{t} A_{t}(x)} \Phi[C] 
,\nonumber \\
A_{\mu\nu}(x) &\rightarrow& A_{\mu\nu}'(x) = A_{\mu\nu}(x) + \partial_{\mu}
\Lambda_{\nu}(x) - \partial_{\nu} \Lambda_{\mu}(x) .
\end{eqnarray}

Then, we perform the old-fashioned dual transformation:
\begin{eqnarray}
\lefteqn{\exp \left\{ i \int d^{10} x -\frac{\varepsilon}{2 \cdot
3!}F_{\mu\nu\lambda}F^{\mu\nu\lambda} \right\} } \nonumber\\
&\propto&\int {\cal D}W_{\mu_{1}\cdots \mu_{7}}(x) \nonumber \\
&&\exp\left\{ i \int d^{10} x \left[ \frac{1}{2 \cdot 7!\varepsilon} 
W_{\mu_{1} \cdots \mu_{7}} W^{\mu_{1} \cdots \mu_{7} } - \frac{1}{7! 3!}
\epsilon^{\mu_{1} \cdots \mu_{10}} W_{\mu_{1} \cdots \mu_{7}} F^{\mu_{8}
\mu_{9} \mu_{10}} \right] \right\}, 
\end{eqnarray}
and integrate out the original variables $A_{\mu\nu}(x)$ as in the previous
section.  The lecture note \cite{Kashikojima} giving the preliminary stage
of this work starts with the action $S^{(0)}$ and proceeds in this way.
Then, we arrive at the dual action $S^{\star} $:
\begin{eqnarray}
S^{\star} &= &\sum_{x} - \frac{1}{8 \cdot 8! (7!)^2 } \frac{1}{\sum_{C(\ni
x)} |\Psi[C]|^2 } F_{\mu_{1} \cdots \mu_{7}} F^{\mu_{1} \cdots \mu_{7}} 
\nonumber \hfill \\
& &+\sum_{S_6} \sum_{x} \bigg| \left( \frac{\delta}{\delta\sigma^{\mu t_{1}
\cdots t_{6}}(x)} - W_{\mu t_{1} \cdots t_{6}}(x) \right) \Psi[S_{6}]
\bigg|^2
\nonumber\\
& &+\cdots . 
\end{eqnarray}
This is the Kalb-Ramond field theory in which the ``6-brane" ($S_{6}$)
interacts with the 7-form field, $W_{\mu_{1} \cdots \mu_{7}}$.
Similarly, we obtain the following correspondences between different brane
theories and their dual brane theories: \begin{equation}
\begin{array}{cccc}
\mbox{0-brane} &\rightarrow & \mbox{7-brane} &(= \mbox{``vortex" connecting
6-branes ``monopole"}) \\
\mbox{1-brane} &\rightarrow & \mbox{6-brane} &(= \mbox{``vortex" connecting
5-branes ``monopole"})  \\
\mbox{2-brane} &\rightarrow & \mbox{5-brane} &(= \mbox{``vortex" connecting
4-branes ``monopole"})  \\
\mbox{3-brane} &\rightarrow & \mbox{4-brane} &(= \mbox{``vortex" connecting
3-branes ``monopole"})
\end{array} \label{dual correspondence}
\end{equation}

A similar study of duality for D-branes is recently carried out in
~\cite{Aurilia}. 

In the above, the present terminology of  ``p-brane" is used to represent
the p-dimensionally extended object in the old days.  The membrane and the
more extended objects (generally p-branes) are now inevitable ingredients,
but have not been popular untill 5 years ago.  Curiously they have the longer
history than the string, since the membranic model of muon (the first
excited state of the elastic ball on which the electric charge is
distributed) by P. A. M. Dirac in 1962 ~\cite{Dirac}. Subsequetly, the
bosonic membrane (p=2) was studied in ~\cite{Collins-Tucker}, and the
bosonic manbrane as well as the general p-brane was studied in
~\cite{Sugamoto}.  As for the spinning membrane, it was first formulated in
~\cite{Howe-Tucker}.  Hereafter, the supermembrane theory is formulated as a
matrix model ~\cite{Hoppe} and it becomes very popular now as a candidate of
the M-theory.

So far we have given the rather naive study of the old fashioned duality
for D-branes.  It is, however, better to examine more carefully each step
of the dual transformation, when we treat the non-local objects such as the
string field functional $\Phi[C]$ and the path integration over the
functionals.  This erases the complexity existing in the preliminary
version of this work, that is  the admixture of the local and the non-local
objects, and may give a clear demonstration of the dual transformation in
string theory.  For this purpose, useful reference is the paper by Marshall
and Ramond~\cite{Marshall-Ramond} even now. 

Let a closed curve $C$ be parmetrized by $X^{\mu}(\lambda)$ with $ 0 \le
\lambda < 2\pi $.  It has the following normal mode expansion,
\begin{equation}
X^{\mu} (\lambda)=\sum_n x^{\mu}_n f_n(\lambda),
\end{equation}
where the normal modes $\{f_n(\lambda)\}$ satisfy the orthonormality and
the completeness conditions:
\begin{equation}
\int d\lambda\, f_m(\lambda)f_n(\lambda)=\delta_{mn}, ~~~\mbox{and}~~~\quad
\sum_n f_n(\lambda)f_n(\lambda')=\delta(\lambda-\lambda').
\end{equation}
Therefore, the string field functional $\Phi [C]$ is a function of infinite
number of variables $\{ x^{\mu}_{n} \}$. Then, the functional derivative
giving the deformation of the curve $C$, 
\begin{equation}
\frac{\delta}{\delta\sigma^{t\mu} (x) } \Phi [C]
\end{equation}
can be defined reparametrization invariant way, by using the orthonormal
expansion:
\begin{equation}
{1\over\sqrt{-({dX^\mu / d\lambda})^2}} {\delta\over\delta X^\mu(\lambda)}
\Phi [X^{\mu}(\lambda)], \label{reparametrization invariant functional
derivative}
\end{equation}
where
\begin{equation}
{\delta\Phi\over\delta X^\mu(\lambda)}=
\sum_n f_n(\lambda){\partial\Phi\over\partial x^\mu_n}.
\end{equation}
If the parametrization $X^{\mu}(\lambda)$ of the curve $C$ in terms of
$\lambda$ is changed to $Y^{\mu}(\rho)$ in terms of $\rho$, defined by
$\lambda = g(\rho) $, the coefficients of the nomal mode expansion is
changed accordingly from $\{x^{\mu}_n \}$ to $\{y^{\mu}_n \}$.  Then, we
have
\begin{equation}
y^\nu_n=\sum_m x^\nu_m\int d\rho\,f_n(\rho)f_m(g(\rho)). 
\end{equation}
From this we can understand that the functional derivative given in
Eq.(\ref{reparametrization invariant functional derivative}) is
reparametrization invariant. Similarly, the reparametrizarion invariant
line integral along the curve $C$ is given by $\int d\lambda\sqrt{-({dX^\mu
/ d\lambda})^2}$.  The unit vector $t^{\mu}$ tangential to the curve $C$
reads $t_\mu(\lambda)\equiv {X'_\mu(\lambda)/\sqrt{-X'^2(\lambda)}}$,
satisfying $t^2=-1$, where $X'^{\mu}(\lambda)$ means $dX^{
\mu}/d\lambda$ as usual.

The important issue is how we perform path integration over the string
field functional $\Phi [C]$.  To solve this problem, we have to know the
way to sum up all the possible shapes of curves $C$.  The integration
measure can be read from the metric (or the distance) of the integration
variables.  The distance $ds$ between two closed curves $C$ and $C+ \delta
C$ can be defined as the minimum area of the membrane connecting $C$ and
$C+ \delta C$ as its boundaries. Since
\begin{equation}
ds^2=\int d\lambda\sqrt{-X'^2}\bigl[
-(\delta X^\mu)^2-(t_\mu\delta X^\mu)^2\bigr] =\sum_{m,n} \sum_{\mu,\nu}
g_{\mu\nu}^{mn}\delta x_m^\mu \delta x_n^\nu ,
\end{equation}
the metric in the space of closed curves is given by
\begin{equation}
g_{\mu\nu}^{mn}\equiv
\int d\lambda\sqrt{-X'^2}[-g_{\mu\nu}-t_\mu t_\nu] f_m(\lambda)f_n(\lambda).
\end{equation}
Therefore, the sum of curves can be done by 
\begin{equation}
\sum_{C} \equiv
\int {\cal D}X \equiv\int \prod_{\mu,n} dx_n^\mu\sqrt{g}\qquad {\rm
with}\quad g\equiv \det g_{\mu\nu}^{mn}.
\end{equation} 

After these preparations, we can write down definitely the action $S$ which
we start with, as follows
\begin{eqnarray}
S=&& \int {\cal D} X^{\mu}(\lambda)
\int d\lambda \sqrt{-X'^2}
\biggl [ ~
\biggl|\biggl({1\over\sqrt{-X'^2}} {\delta\over\delta X^\mu(\lambda)}
- it^\nu(\lambda) A_{\nu\mu}[C,X(\lambda)]\biggr)\Phi[C]\biggr|^2 \nonumber \\
+&& \frac{1}{12} F_{\mu\nu\rho}[C, X(\lambda)] F^{\mu\nu\rho} [C,
X(\lambda)] - V(|\Phi[C]|^2) ~\biggr ], 
\end{eqnarray}
where $F_{\mu\nu\rho}[C, X(\lambda)]$ is the field strength of the
Kalb-Ramond field $A_{\mu\nu}[C,X(\lambda)]$, defined by 
\begin{equation}
F_{\mu\nu\rho}[C,X(\lambda)]\equiv
{1\over\sqrt{-X'^2}}
{\delta A_{\nu\rho}[C,X(\lambda)]\over\delta X^\mu(\lambda)} +
\mbox{(cyclic permutations)}.
\end{equation}

It should be noted that the Kalb-Ramond field, or the gauge field of the
string, is taken to be a non-local one defined on the curve $C$ and the
point $X^{\mu}(\lambda)$.  This non-locality is necessary to carry out the
dual transformation consistently. The action is shown to be invariant under
the Kalb-Ramond transformation, or the gauge transformation of the 2-form
field, namely,
\begin{eqnarray}
\Phi[C] &\rightarrow&\exp (i\Lambda[C]) \Phi[C],\nonumber \\
A_{\mu\nu}[C,X(\lambda)] &\rightarrow&
A_{\mu\nu}[C,X(\lambda)]
+{t_\mu(\lambda)\over\sqrt{-X'^2}}
{\delta\Lambda[C]\over\delta X^\nu(\lambda)} -{t_\nu(\lambda)\over\sqrt{-X'^2}}
{\delta\Lambda[C]\over\delta X^\mu(\lambda)}.
\end{eqnarray} 

It is also possible to choose the simpler action $S'$ rahter than $S$,
\begin{eqnarray}
S'=&& \int {\cal D} X^{\mu}(\lambda)
\int d\lambda\sqrt{-X'^2}
\biggl [ ~
\biggl|\biggl({1\over\sqrt{-X'^2}} {\delta\over\delta X^\mu(\lambda)}
-i A_{\mu}[C,X(\lambda)]\biggr)\Phi[C]\biggr|^2 \nonumber \\ 
&&- \frac{1}{4} F_{\mu\nu}[C,X(\lambda)] F^{\mu\nu}[C,X(\lambda)] -
V(|\Phi[C]|^2)~\biggr ],
\end{eqnarray}
where the field strength of the non-local gauge field is given by 
\begin{equation}
F_{\mu\nu}[C,X(\lambda)]\equiv
{1\over\sqrt{-X'^2}}
{\delta A_{\nu}[C,X(\lambda)]\over\delta X^\mu(\lambda)}
-(\mu\longleftrightarrow\nu),
\end{equation}
and the action is invariant under the following gauge transformation 
\begin{equation}
A_{\mu}[C,X(\lambda)]\rightarrow
A_{\mu}[C,X(\lambda)]+{1\over\sqrt{-X'^2}} {\delta\Lambda[C]\over\delta
X^\mu(\lambda)}.
\end{equation} 

We examine in detail the dual transformation of the former action $S$ in
the following.  For the kinetic term of the Kalb-Ramond fields in  $S$, we
apply the following dual transformation:
\begin{eqnarray}
&&\exp\biggl[ \frac{i}{12} \int {\cal D}X\int d\lambda\sqrt{-X'^2}
F_{\mu\nu\rho}F^{\mu\nu\rho}[C,X(\lambda)] \biggr] \nonumber \\
&&=\int {\cal D}W^{\mu_1 \mu_2, \dots, \mu_7}
\exp\biggl[ \frac{i}{12} \int {\cal D}X^{\mu}(\lambda)\int
d\lambda\sqrt{-X'^2} \bigl\{- \left( {\tilde W}^{\mu\nu\rho} \right)^2 +2
{\tilde W}^{\mu\nu\rho} F_{\mu\nu\rho}\bigr\} \biggr].  \label{DT of 2-form
field}
\end{eqnarray}
Here, the path integration over $X^{\mu}(\lambda)$ is defined in the above
in terms of the metric 
\begin{equation}
\int {\cal D} X^{\mu}(\lambda) \equiv 
\int d\xi^{
\mu}_{n} \sqrt{g(\xi)}.
\end{equation}
Writing the compex string field functional as $\Phi[C]=
|\Phi[C]|e^{i\chi[C]}$, and  using the integration by parts, the relevant
terms to $A_{\mu\nu}[C, X(\lambda)]$ read
\begin{eqnarray}
\int {\cal D} X^{\mu}(\lambda) \int d\lambda \sqrt{ -X'^{2}} &&\biggl[
(t^{\mu}(\lambda)A_{\nu\mu}) (t^{\rho}(\lambda)A_{\rho}^{\mu})
|\Phi[C]|^{2} - \frac{1}{2\sqrt{g}} \frac{1}{\sqrt{-X'^{2}}}
\frac{\delta}{\delta X^{\mu}}(\sqrt{g}\tilde{W}^{\mu\nu\rho}) A_{\nu\rho}
\nonumber \\
&&-2 |\Phi[C]|^{2} \frac{1}{\sqrt{-X'^{2}}} \frac{\delta \chi[C]}{\delta
X^{\mu}} t^{\nu} A_{\nu\mu} \biggr ].
\end{eqnarray}
Here, we decompose the vector index into the tangential direction $t^{\mu}$
along $C$, and the transverse directions to it.  Defining $B_{\nu}\equiv
t^{\mu}A_{\mu\nu}$ and $V_{\nu\rho}\equiv t^{\mu}{\tilde W}_{\mu\nu\rho}$,
the decomposition of $A_{\mu\nu}$, and ${\tilde W}_{\mu\nu\lambda}$ is
given by
\begin{eqnarray}
A_{\mu\nu}&=&-(t_{\mu}B_{\nu}-t_{\nu}B_{\mu}) + A_{\mu\nu}^{T},\nonumber \\
{\tilde W}_{\mu\nu\lambda}&=&-(t_{\mu}V_{\nu\lambda}+ \mbox{cyclic perm.})
+ {\tilde W}_{\mu\nu\rho}^{T},
\end{eqnarray}
where we have $t^{\mu}B_{\mu}=0$, $t^{\mu}A_{\mu\nu}^{T}=0$,
$t^{\mu}{\tilde V}_{\mu\nu}=0$, and $t^{\mu}{\tilde
W}_{\mu\nu\lambda}^{T}=0$.
Then, the path integration over $A_{\mu\nu}^{T}$ gives a constraint 
\begin{equation}
{\delta\over\delta X^\rho(\lambda)}
\biggl(\sqrt{g}{\tilde W}^{T\mu\nu\rho}\biggr)=0, 
\end{equation} 
while the path integration over $B_{\mu}$ is the Gaussian integration.

Accordingly, the following dual action $S^{\star}$ is derived, starting
with the original string action $S$,
\begin{eqnarray}
S^{\star}=&&\int {\cal D} X^{\mu}(\lambda) \int d\lambda \sqrt{-X'^{2}} 
\biggl[
-{1\over 4g|\Phi[C]|^2}\biggl\{{1\over\sqrt{-X'^2}} {\delta \over \delta
X^{\mu}}\biggl(
\sqrt{g}{\tilde V}^{\mu\nu}\biggr)\biggr\}^2 +{1\over4} {\tilde
V}_{\mu\nu}^2 \nonumber \\
&&-{1\over12} ({\tilde W}^T_{\mu\nu\rho})^2 
-{1\over\sqrt{-X'^2}}
{\delta\over\delta X^{\mu}}\biggl({1\over\sqrt{-X'^2}}
{\delta\chi[C]\over\delta X^\nu}\biggr){\tilde V}^{\mu\nu}
+\biggl({1\over\sqrt{-X'^2}}
{\delta|\Phi[C]| \over \delta  X^{\mu}}\biggr)^2  \biggr]. \label{dual
action of S}
\end{eqnarray}
Similarly, from the second action of $S'$, we can derive the following dual
action:
\begin{eqnarray}
S'^{\star}=&&\int {\cal D} X^{\mu}(\lambda) \int d\lambda \sqrt{-X'^{2}} 
\biggl [
-{1\over 4g|\Phi[C]|^2}\biggl\{{1\over\sqrt{-X'^2}} {\delta \over \delta
X^{\mu}}\biggl(
\sqrt{g}{\tilde W}^{\mu\nu}\biggr)\biggr\}^2 +{1 \over 4}{\tilde
W}_{\mu\nu}^2 \nonumber\\
&&-{1 \over \sqrt{-X'^2}}
{\delta \over \delta X^{\mu}}\biggl({1 \over \sqrt{-X'^2}} {\delta\chi[C]
\over \delta X^{\nu}}\biggr){\tilde W}^{\mu\nu} +\biggl({1 \over
\sqrt{-X'^2}}
{\delta|\Phi[C]| \over \delta X^{\mu}}\biggr)^2 \biggr]. \label{dual action
of S'}
\end{eqnarray}
These two dual actions are equivalent, since they are related to each other:
\begin{equation}
S'^{\star}=S^{\star}
+ \int {\cal D} X^{\mu}(\lambda) \int d\lambda \sqrt{-X'^{2}} {1\over
4g|\Phi[C]|^2}\biggl\{{1\over\sqrt{-X'^2}} {\delta \over \delta
X^{\mu}}\biggl(
\sqrt{g}C^\mu\biggr)\biggr\}^2
-{1 \over 2}C_\mu^2, 
\end{equation}
where $C_{\mu}\equiv t^\mu \tilde{W}_{\mu\nu}$, and
$\tilde{W}_{\mu\nu}\equiv -(t_\mu C_\nu-t_\nu C_\mu)+ \tilde{V}_{\mu\nu}$
with $t^\mu C_\mu=0$ and $t^\mu\tilde{V}_{\mu\nu}=0$.

In the dual actions in Eqs. [\ref {dual action of S}][\ref {dual action of
S'}], the ``vorticity source" $\omega_7(S_7)$ is given by
\begin{eqnarray}
&&\omega_{\mu_1\mu_2, \dots, \mu_7}(S_7) \equiv \frac{1}{4
\pi} \epsilon_{\mu_1\mu_2 \dots \mu_7 \nu_1\nu_2\rho} ~t^{\rho} \nonumber \\
&&\times\left( {1\over\sqrt{-X'^2}} {\delta \over \delta X^{\nu_1}}
{1\over\sqrt{-X'^2}} {\delta\over\delta X^{\nu_2}} - {1\over\sqrt{-X'^2}}
{\delta \over \delta X^{\nu_2}} {1\over\sqrt{-X'^2}} {\delta\over\delta
X^{\nu_1}} \right) \chi[C] .
\label{vorticity by string} \end{eqnarray} 
To extract the ``vorticity source" from the singularity of the phase
$\chi[C]$ of the wave functional, is not a simple task. The task means to
understand how the various topological quantum numbers are defined using
the non-local field, such as the string field $\Phi[C]$ or $\chi[C]$.  More
generally, the problem is how are the topological quantum numbers defined
with the p-branes' wave functional?  This is an interesting mathematical
problem probably to be solved in the non-commutative
geometry~\cite{Connes}.  It is because, the string and the p-brane theories
in general give the non-commutativity in space-time, so that the geometry
behind these topological quantum numbers should be the non-commutative
geometry. 

Here, we study a simple case in which the mean value $<\chi[C]>$ of
$\chi[C]$ changes $2\pi n$ while we deform the curve $C$ transversally and
return to the original shape of $C$.  The homotopy of this deformation can
be given as 
$X^{\mu}(\lambda, s_1)=\sum_{m} f_{m}(\lambda)x^{\mu}_{m}(s_1)$
with the parameter
$s_1 (0\le s_1\le 2\pi)$.  At $s_1=0$ and $\pi$, $X^{\mu}(\lambda, s_1)$ gives
 the
curve $C(X^{\mu}(\lambda))$. Then, we have
\begin{eqnarray}
 2\pi n(\int d\lambda \sqrt{-X'^{2}})
 &=&\int d\lambda \sqrt{-X'^{2}} \oint
\frac{\partial\chi}{\partial x^{\mu}_{m}(s_1)}~dx^{\mu}_{m}(s_1) \nonumber\\
&=&\int
d\lambda \sqrt{-X'^{2}}\int\int ds_1 ds_2 \frac{\partial^{2} \chi}{\partial
x^{\mu}_{m}\partial x^{\nu}_{n}} \frac{\partial (x^{\mu}_{m},
x^{\nu}_{n})}{\partial(s_1, s_2)}. \hfill
\end{eqnarray}
This means
\begin{equation}
\frac{\partial^{2}\chi}{\partial x^{\mu}_{m}\partial x^{\nu}_{n}}-
\frac{\partial^{2}\chi}{\partial x^{\nu}_{n}\partial x^{\mu}_{m}}= 4\pi
n(\int d\lambda \sqrt{-X'^{2}}){1\over\sqrt{-X'^{2}}} \frac{\partial(s_1, s
_2)}{\partial (x^{\mu}_{m}, x^{\nu}_{n})}\delta(s_1-s^{*}_1)
\delta(s_2-s^{*}_2) \delta(\lambda-\lambda^{*}).
\end{equation}
If we add 7 parameters $(\sigma_1, \cdots, \sigma_7)$ in addition to
$(\lambda, s_1, s_2)$, then the whole 10
dimensional space can be parametrized by $x^{\mu}=Y^{\mu}(\lambda, s_1,
s_2, \sigma_1, \cdots, \sigma_7)$. The singularity existing at $(\lambda,
s_1, s_2)=(\lambda^{*}, s^{*}_1, s^{*}_2)$ determines $S_7$ (the world
volume of the 6-brane) by $\{Y^{\mu}(\sigma_1, \cdots, \sigma_7)\}$, on
which the vorticity takes the non-vanishing value.
Now, the vorticity source can be written as
\begin{eqnarray}
&&\omega_{\mu_1\mu_2, \dots, \mu_7}(S_7) \nonumber \\
&&= -n (\int d\lambda \sqrt{-X'^{2}}) \int d\sigma_1 d\sigma_2 \cdots
d\sigma_7 ~ {\sqrt{-X'^{2}}}^{-2} \frac{\partial (Y_{\mu_1}, Y_{\mu_2}, \cdots,
Y_{\mu_7})} {\partial (\sigma_1, \sigma_2, \cdots, \sigma_7)} \nonumber \\
&&\times \delta^{(10)} (x^{\mu}-Y^{\mu}(\sigma_1, \sigma_2, \cdots,
\sigma_7)). \hfill
\end{eqnarray}
Then, the world volume $S_7$ of the 6-brane couples minimally to the
Kalb-Ramond 7-form field, or  $\tilde{W}_{\mu\nu\rho}t^{\rho}$, where
$\tilde{W}_{\mu\nu\rho}$ and $A_{\mu\nu}$ are dually related. 

Our formulation is adequate for extracting the vortex rather than the
monopole, so that the derived 6-brane is considered to be the bare
``magnetic vortex", connecting the pair of ``magnetic monopoles" on the both
ends.  Therefore, the ``magnetic monopole" is the 5-brane which is the dual
object of the ``fundamental string" of the 1 brane.  The other
correspondences are given in Eq.(\ref{dual correspondence}).


\section{Old-Fashioned Duality for Superstrings} 
We know that the place where the duality plays the powerful role is the
superstring theories, so that our next step should be the application of
the old-fashioned duality for superstrings and see how it works. However,
we have to restrict ourselves only to a small step towards this project, by
examining the old fashioned dual transformation in the ``field theories of
the spining strings".

In the spinning string of Neveu-Schwarz and Ramond, ten spins are attached
on the curve $C$, which are reprensented by the two sets ($i=1, 2$) of ten
2-dimensional spinors $\psi_{(i)}^{\mu}(\lambda) (\mu=0-9,~~\mbox{and}~~
i=1, 2)$.  They satisfy the  commutation relations,
\begin{equation}
\{ \psi^{\mu}_{(i)} (\lambda), \psi^{\nu}_{(j)}(\lambda') \} = 
G^{\mu\nu}\delta_{ij}\delta(\lambda-\lambda'). \label{CR's for psi}
\end{equation}
Following~\cite{Marshall-Ramond}, the mode expansion of $\psi_{(i)}^{\mu}$
for the Neveu-Schwarz (NS) sector is given by
\begin{equation}
\psi^{\mu}_{(i)} (\lambda)_{(NS)} \equiv b^{\mu}_{(i)}(\lambda) = \sum_{k:
\mbox{\footnotesize half-integer}} b^{\mu}_{(i) k} f_{k} (\lambda),
\end{equation} while for the Ramond (R) sector,
\begin{eqnarray}
\psi^{\mu}_{(1)} (\lambda)_{(R)} &\equiv&
\Gamma^{\mu}_{(1)}(\lambda)/\sqrt{2} = \gamma^{\mu}/\sqrt{2} + \gamma_{11}
\sum_{n: \mbox{\footnotesize integer} \ne 0} d^{\mu}_{(1) n} f_{n}
(\lambda), \nonumber \\
\psi^{\mu}_{(2)} (\lambda)_{(R)} &\equiv&
\Gamma^{\mu}_{(2)}(\lambda)/\sqrt{2} =\gamma_{11} \sum_{n:
\mbox{\footnotesize integer including 0}} d^{\mu}_{(2)n} f_{n} (\lambda).
\end{eqnarray}
Here, $\psi^{\mu}_{(i)} (\lambda)$ is related to the right-moving mode
$\psi(\tau-\sigma)$ and the left-moving mode $\tilde{\psi} (\tau+\sigma)$
as follows~\cite{Polyakov}:
\begin{eqnarray}
\psi^{\mu}_{(1)}&=&\tilde{\psi^{\mu}}+\psi^{\mu}, \nonumber \\
\psi^{\mu}_{(2)}&=&\tilde{\psi^{\mu}}-\psi^{\mu}.
\end{eqnarray}
The 10 dimensional spinor structure is naturally build in  with the help of
the $\gamma^{\mu}$ matrices which are lifted to the position dependent
fields $\Gamma^{\mu}_{(1)} (\lambda)$ on the curve $C$ (See
~\cite{Marshall-Ramond}).
From the commutation relation in Eq. (\ref{CR's for psi}), we understand
the product $[- X'^{2}]^{-1/4} \psi^{\mu}(\lambda) \equiv
\hat{\psi}^{\mu}(\lambda)$ is reparametrization invariant. Similarly, the
reparametrization invariant fields $\hat{\Gamma}^{\mu}(\lambda)$ and
$\hat{b}^{\mu}(\lambda)$ are introduced by multiplying $[- X'^{2}]^{-1/4}$.
The string field functional of the spinning string depends on the shape of
the curve $C$, as well as the spins' configuration.  Therefore, the string
field functional in the NS sector is bosonic and is given by
$\Phi[X^{\mu}(\lambda), ~b^{\mu}_{1, 2}(\lambda)]$, while the string field
functional in the R sector is a 10 dimensional spinor
$\Psi[X^{\mu}(\lambda), ~d^{\mu}_{1, 2}(\lambda)]$.

Now, we can write down the field theoretical superstring actions.  Bosonic
part of the action $S_{b}$ is common for Type IIA and Type IIB, namely
\begin{eqnarray}
\lefteqn{S_{b} = \int {\cal D} X^{\mu}(\lambda) \int {\cal D} 
b^{\mu}_{(1)}(\lambda){\cal D}  b^{\mu}_{(2)}(\lambda) \int d\lambda \sqrt
{-X'^{2}}~\bigg[  \Phi [C, b^{\mu}_{(1, 2)}(\lambda)]^{*} \hat{b}^{0}_{(1)}
(\lambda) } \nonumber \\
&&\times~ \left\{ e^{D[C, X(\lambda)]} E^{\nu}_{\bar {\mu}} [C, X(\lambda)]
\right\} \bigg\{ 
i \hat{b}^{\bar {\mu}}_{(1)}(\lambda)  \bigg( {1\over\sqrt{-X'^2}}
{\delta\over\delta X^\nu(\lambda)}
- it^\rho(\lambda) B_{\rho\nu}[C,X(\lambda)] \nonumber\\
&&+~ (1/8)\left[ \hat{b}_{(1)\bar{\mu}}(\lambda) \hat{b}_{(1)\bar
{\nu}}(\lambda) \right] \Omega^{\bar{\mu} \bar{\nu}}_{\nu} [C, X(\lambda)]
\bigg) +
(1/\alpha' ) \hat{b}_{(2)}^{\bar{\mu}} (\lambda) t_{\nu}(\lambda)
\bigg\}
 \Phi [C, b^{\mu}_{(1, 2)}(\lambda)]
\bigg], \hfill
\end{eqnarray}  
where the dilaton, vierbein, spin connection, and the
2-form field (Kalb-Ramond field) are denoted by $D[C, X(\lambda)], E^{\bar
{\mu}}_{\nu}[C, X(\lambda)], ~\Omega^{\bar {\mu}\bar {\nu}}_{\mu} [C,
X(\lambda)]$, and $B_{\rho\nu}[C, X(\lambda)]$, respectively.  We use the
indices without and with bars for the flat and curved ones, respectively.

The fermionic action depends on the type of the closed string theories,
namely Type IIA or IIB.  For the Type IIB superstring, we have the coupling
of the chiral fermion with the even form fields $A^{(p)}$ ($p=$even) in the
R-R sector, so that the action $S_{f}(\mbox{IIB})$ is given by
\begin{eqnarray}
S_{f}(\mbox{IIB}) & = & \int {\cal D} X^{\mu}(\lambda) \int {\cal D} 
d^{\mu}_{(1)}(\lambda) \int {\cal D}  d^{\mu}_{(2)}(\lambda)
\int d\lambda \sqrt {-X'^{2}} \bigg[ \Psi_{L}[C, d^{\mu}_{(1, 2)}(\lambda)]^{T}
\hat{\Gamma}^{0}_{(1)}(\lambda) \nonumber \\
&& \times~ \bigg\{
i\hat{\Gamma}^{\mu}_{(1)}(\lambda) 
{1\over\sqrt{-X'^2}}
{\delta\over\delta X^\mu(\lambda)}
+ t^\tau(\lambda) \hat{\Gamma}_{\tau (1)}(\lambda) A^{(0)}[C, X(\lambda)]
\nonumber\\
&& +~ t^\tau(\lambda) \hat{\Gamma}^{\mu}_{(1)}(\lambda)
A_{\tau\mu}^{(2)}[C,X(\lambda)] \nonumber \\
&& +~ t^\tau(\lambda) \hat{\Gamma}^{\mu}_{(1)} (\lambda)
\hat{\Gamma}^{\nu}_{(1)} (\lambda) \hat{\Gamma}^{\rho} _{(1)} (\lambda)
A_{\tau\mu\nu\rho}^{(4)}[C, X(\lambda)] \bigg \} 
\Psi_{L} [C, d^{\mu}_{(1, 2)}(\lambda)] \bigg]. \hfill \label{TypeIIB}
\end{eqnarray}
 Here, the product of even number of $\hat{\Gamma}^{\mu}_{(1)}$'s has
appeared between $[\Psi_{L}]^{T}$ and $\Psi_{L}$, so that the theory
becomes chiral, without having the R-handed field functional.  Notice that
the $\hat{\Gamma}^{\mu}_{(2)}$ does not change the chirality of the
fermionic field functional.

As for the Type IIA superstrings, the fermionic field functionals of both
chiralities are introduced, which couple to the odd form fields in the R-R
sector, giving the following fermionic action $S_{f}(\mbox{IIA})$:
\begin{eqnarray}
S_{f}(\mbox{IIA}) &=& \int {\cal D} X^{\mu}(\lambda) \int {\cal D} 
d^{\mu}_{(1)}(\lambda) \int {\cal D}  d^{\mu}_{(2)}(\lambda) 
\int d\lambda \sqrt {-X'^{2}} \bigg[ \Psi[C, d^{\mu}_{(1, 2)}(\lambda)]^{T}
\hat{\Gamma}^{0}_{(1)}(\lambda)  \nonumber \\
&&\times~ \bigg\{
i\hat{\Gamma}^{\mu}_{(1)}(\lambda) {1\over\sqrt{-X'^2}}
{\delta\over\delta X^\mu(\lambda)}
+ t^\tau(\lambda) A_{\tau}^{(1)}[C, X(\lambda)] \nonumber \\
&&+~ t^\tau(\lambda) \hat{\Gamma}^{\mu} _{(1)}(\lambda)
\hat{\Gamma}^{\nu}_{(1)} (\lambda) A_{\tau\mu\nu}^{(3)}[C, X(\lambda)]
\nonumber\\
&&+~ t^\tau(\lambda) \hat{\Gamma}^{\mu}_{(1)} (\lambda)
\hat{\Gamma}^{\nu}_{(1)} (\lambda) \hat{\Gamma}^{\rho} _{(1)} (\lambda)
\hat{\Gamma}^{\sigma} _{(1)} (\lambda) 
A_{\tau\mu\nu\rho\sigma}^{(5)}[C, X(\lambda)] \nonumber \\
&&+~ (1/\alpha') \hat{\Gamma}^{\mu}_{(2)}(\lambda) t_{\mu}(\lambda)
\bigg\} \Psi[C, d^{\mu}_{(1, 2)}(\lambda)] \bigg].  \label{TypeIIA}
\end{eqnarray} 
In this case, the coupling of the odd form fields $A^{(p)}$ ($p=$odd) in
the R-R sector changes the chirality of the field functional.

The transformation property of the R-R p-form field $A^{(p)}$ for $p\ge 2$
are summarized as follows
\begin{eqnarray}
\Psi[C, d^{\mu}(\lambda)] &\rightarrow&\exp \left( i \hat{\Gamma}^{ [
\mu_1, \cdots, \mu_{p-2} ] } \Lambda_{\mu_1, \cdots, \mu_{p-2}}[C] \right)
\Psi[C, d^{\mu}(\lambda)], \\
A^{(p)}_{\mu_1\mu_2, \cdots, \mu_p}&\rightarrow& A^{(p)}_{\mu_1\mu_2,
\cdots, \mu_{p}} + \frac{t_{\mu_1}}{\sqrt{-X'^2}}
\frac{\delta\Lambda_{\mu_3, \cdots, \mu_{p}}}{\delta X^{\mu_2}} +
(\mbox{cyclic perm.}). \end{eqnarray}
The transformation property of the 0- and 1-form fields of the R-R sector
is not clear in this formulation, but  the reason why even and odd forms of
the R-R sector couple to the fermionic field functional of Type IIB and IIA
superstrings, respectively, is well understood. Transformation property of
the NS-NS sector is not special. The local Lorentz transformation is
generated by 
\begin{equation}
M^{\bar{\mu} \bar{\nu}}(\lambda)=\frac{1}{4}~[\Gamma^{\bar{\mu}}(\lambda),
\Gamma^{\bar{\nu}}(\lambda)], 
\end{equation}
while the general coordinate transformation is generated by
\begin{equation}
T'^{\mu'\nu'\cdots}[C, X'(\lambda)]=\left( \frac{\sqrt{-X'^2(\lambda)}
\delta X'^{\mu'}(\lambda)}{\sqrt{-X^2 (\lambda)}\delta X^{\mu} (\lambda)}
\right) \left( \frac{\sqrt{-X'^2 (\lambda)} \delta X'^{\nu'}
(\lambda)}{\sqrt{-X^2 (\lambda)} \delta X^{\nu} (\lambda)} \right) \cdots
T^{\mu\nu\cdots}[C, X(\lambda)],
\end{equation}
where the general cordiante transformation is naturally modified
$\lambda$-dependently:
\begin{equation}
\frac{\delta X'^{\nu} (\lambda)}{\delta X^{\mu} (\lambda)}~=~\sum_{n,m}~~
\frac{\partial x'^{\nu}_{n}}{\partial x^{\mu}_{n}}
f_{n}(\lambda)f_{m}(\lambda).
\end{equation}
Now, starting with these superstring actions, we examine the old-fashioned
dual transformation for them.  In addition to the bosonic and fermionic
actions, we have the kinetic terms of the NS-NS fields and of the R-R
p-from fields $A_p~(p=0, 2, 4)$.  The kinetic action of the R-R fields
reads
\begin{eqnarray}
S_0(\mbox{IIB}) &=& \int {\cal D} X^{\mu}(\lambda)
\int d\lambda \sqrt{-X'^2}
\bigg[ (1/2)F_{(0)\mu}[C, X(\lambda)] F_{(0)}^{\mu} [C, X(\lambda)]
\nonumber \\ 
&&+~(1/12)F_{(2)\mu\nu\rho}[C, X(\lambda)] F_{(2)}^{\mu\nu\rho} [C,
X(\lambda)] \nonumber\\
&&+~ (1/240)F_{(4)\mu\nu\rho\sigma\tau}[C, X(\lambda)]
F_{(4)}^{\mu\nu\rho\sigma\tau} [C, X(\lambda)] ~\bigg]. \hfill
\end{eqnarray}
Using the similar equation to Eq. (\ref{DT of 2-form field}), we can
replace the R-R fields to their dual fields.  For example for the R-R
2-form field, the path integration over the field $A_{(2)}$ gives  the
following constraint:
\begin{eqnarray}
&&{1\over\sqrt{g}} \frac{1}{\sqrt{-X'^{2}}} \frac{\delta}{\delta
X^{\mu}}(\sqrt{g}\tilde{W}_{(2)}^{\mu\nu\rho}) \nonumber \\
&&=2\int {\cal D}  d^{\mu}_{(1)}(\lambda) \int {\cal D} 
d^{\mu}_{(2)}(\lambda) \Psi_{L}[C, d_{(1,
2)}]^{T}\hat{\Gamma}^{0}_{(1)}\hat{\Gamma}^{\rho}_{(1)} t^{\nu}\Psi_{L}[C,
d_{(1, 2)}] \equiv J^{\nu\rho}[C, X(\lambda)]. \hfill
\end{eqnarray}
This constraint can be solved as
\begin{eqnarray}
\sqrt{g}\tilde{W}_{(2)\mu\nu\rho}&=&{1\over6!}\epsilon_{\mu\nu\rho\tau\sigma_1,\cdots,\sigma_6} \frac{1}{\sqrt{-X'^{2}}} \frac{\delta}{\delta
X_{\tau}} B_{(6)}^{\sigma_1, \cdots,\sigma_6} \nonumber \\
&&+\int {\cal D}Y^{\nu}(\rho)\frac{1}{\sqrt{-X'^{2}}} \frac{\delta}{\delta
X_{\tau}} D [C\{ X^{\mu}(\lambda) \}, C'\{ Y^{\nu}(\rho) \}] \sqrt{g}
J^{\nu\rho}[C'].\hfill
\end{eqnarray}
Here, the first term in the r.h.s. is the general solution in the case of
$J^{\nu\rho}=0$, and the second term is given, using the propagator $D[C,
C']$ of the bosonic string field theory.  Namely, $D[C, C']$ is defined by
\begin{equation}
\left( \frac{1}{\sqrt{-X'^{2}}} \frac{\delta}{\delta X_{\mu}} \right)^2
D[C\{ X^{\mu}(\lambda) \}, C'\{ Y^{\nu}(\rho) \}] = \delta[C\{
X^{\mu}(\lambda) \}, C'\{ Y^{\nu}(\rho) \}].
\end{equation}
Now, the relevant terms of the dual action become
\begin{eqnarray}
S^{\star}=&&\int {\cal D} X^{\mu}(\lambda) \int d\lambda \sqrt{-X'^{2}}\int
{\cal D}  d^{\mu}_{(1)}(\lambda) \int {\cal D}  d^{\mu}_{(2)}(\lambda)
\nonumber \\
&&\times \Psi_{L}[C, d^{\mu}_{(1, 2)}(\lambda)]^{T}
\hat{\Gamma}^{0}_{(1)}(\lambda) 
i\hat{\Gamma}^{\mu}_{(1)}(\lambda) 
{1\over\sqrt{-X'^2}}
{\delta\over\delta X^\mu(\lambda)}
\Psi_{L} [C, d^{\mu}_{(1, 2)}(\lambda)] \nonumber \\
&&+ \int {\cal D} X^{\mu}(\lambda) \int d\lambda \sqrt{-X'^{2}} 
\biggl[-\frac{1}{12}\left( \frac{1}{\sqrt{-X'^{2}}} \frac{\delta}{\delta
X_{\mu}}\Phi^{\nu\rho}[C] + \tilde{H}^{\mu\nu\rho}\right)^{2}
+ \dots . \hfill \label{IIBdual} 
\end{eqnarray}
  Here, the $\tilde{H}^{\mu\nu\rho}$ is the dual tensor of the field
strength of $B_{(6)}$,

\begin{equation}
\tilde{H}_{\mu\nu\rho}=\frac{1}{6!} \epsilon_{\mu\nu\rho\sigma_1\sigma_2,
\cdots, \sigma_7}\frac{1}{\sqrt{-X'^{2}}} \frac{\delta}{\delta X_{\sigma_1}}
B_{(6)}^{\sigma_2, \cdots,\sigma_6},
\end{equation}
and
\begin{equation}
\Phi^{\nu\rho}[C]=\int {\cal D}Y^{\nu}(\rho) D [C\{ X^{\mu}(\lambda) \},
C'\{ Y^{\nu}(\rho) \}] \sqrt{g} J^{\nu\rho}[C']. \hfill
\end{equation}
After the dual transformation of the Type IIB superstring is performed, the
obtained action in Eq. (60) includes the free fermionic and 4-fermionic
terms as well as the coupling of the fermionic current to the field
$H_{(3)}$ which is dually related to the RR 2-form field $A_{(2)}$.

From the above consideration, a kind of ``bosonization" should work for the
NS-NS and R-R field functionals, $\Phi[C]$ and $\Psi_{L}[C]$. Namely the
$\Phi[C]^{\star}$ in the dual 
action is the ``bosonization" of the $\Psi_{L}[C]$ in the original action,
and the $\Psi_{L}[C]^{\star}$ in the dual action is the ``fermionization" of
the $\Phi[C]$ in the original action. This conjecture is by no means
unrealistic, since the field functionals are defined on the closed curve
$C$ and are essentially the 2 dimensional objects.  If this kind of
``bosonization" happens, then the free fermionic term and the 4-fermionic
curent-current interaction term for the R-R field functional appearing in
the dual action (60) becomes the free bosonic action of the NS-NS field
functional, so that the dual action approaches the NS-NS action of the Type
IIB superstring. The self-dualty of the Type IIB superstring may be proved
in this way within the framework of the old fashioned duality.

The old fashioned dual transformation transforms the gauge theory to the
hydrodynamics of Kalb-Ramond and Nambu~\cite{K-R} coupled to the extended
objects of p-branes.  Therefore, the same technique is applicable to the
problem of swimming of microorganisms in the fluid.  Since the envelope of
the microorganisms can be considered as the closed string or membrane which
couples naturally to the outside viscous fluid~\cite{microorganisms}. The
study in this direction is another interesting application of the
duality~\cite{duality and microorganisms}.

\section*{Acknowledgment}
We are grateful to Shin'ichi Nojiri and Sergei Odintsov for their
continuous encouragements extended to us.  Without their support we could
not finish this work.



\end{document}